\pdfoutput=1
\documentclass[aps,reprint,amsmath,amssymb,superscriptaddress]{revtex4-1}
\usepackage{bm}
\usepackage{graphicx}
\usepackage{braket}
\usepackage[colorlinks=true,linkcolor=red, citecolor=blue, urlcolor=blue, bookmarks]{hyperref}
\usepackage{mathrsfs}
\usepackage{times}
\usepackage[normalem]{ulem}
\usepackage{bbm}

\newcommand{\hpa}{h_{\parallel}}
\newcommand{\hpe}{h_{\perp}}
\newcommand{\I}{\text{i}}

\newcommand{\dd}{{\rm d}}
\newcommand{\be}{\begin{equation}}
\newcommand{\ee}{\end{equation}}

\newcommand{\cut}[1]{}

\begin{document}
\newcommand{\titleinfo}{
Entanglement dynamics in confining spin chains
}

\title{\titleinfo}

\author{Stefano Scopa}
\affiliation{SISSA and INFN, via Bonomea 265, 34136 Trieste, Italy}

\author{Pasquale Calabrese}
\affiliation{SISSA and INFN, via Bonomea 265, 34136 Trieste, Italy}
\affiliation{
International Centre for Theoretical Physics (ICTP), I-34151, Trieste, Italy}

\author{Alvise Bastianello}
\affiliation{Department of Physics, Technical University of Munich, 85748 Garching, Germany}
\affiliation{Institute for Advanced Study,  85748 Garching, Germany}
\affiliation{Munich Center for Quantum Science and Technology (MCQST), Schellingstr. 4, D-80799 M{\"u}nchen, Germany}

\begin{abstract}
The confinement of elementary excitations induces distinctive features in the non-equilibrium quench dynamics. 
One of the most remarkable is the suppression of entanglement entropy which in several instances turns out to oscillate rather than grow indefinitely. While the qualitative physical origin of this behavior is clear, till now no quantitative understanding away from the field theory limit was available.
Here we investigate this problem in the weak quench limit, when mesons are excited at rest, hindering entropy growth and exhibiting persistent oscillations.
We provide analytical predictions of the entire entanglement dynamics based on a Gaussian approximation of the many-body state, 
which captures numerical data with great accuracy and is further simplified to a semiclassical quasiparticle picture in the regime of weak confinement. Our methods are valid in general and we apply explicitly to two prototypical models: the Ising chain in a tilted field and the experimentally relevant long-range Ising model.
\end{abstract}

\maketitle
\section{Introduction}
The relentless advances in cold atom experiments made available extremely tunable platforms where synthetic phases of matter can be engineered with high precision.
Quantum simulators \cite{feynman2018simulating,Georgescu2014quantum} are on the verge of replacing classical devices in the challenge of understanding and faithfully describing many-body and strongly interacting quantum systems, promising a pathway to simulations that are far beyond the reach of classical computers. 
A prominent example is the physics of strongly coupled gauge theories routinely probed at large hadron colliders.
Even though the promised land of an accurate quantum simulator for such complex systems is getting closer to a fast pace, it nowadays remains still low on the horizon. Recent times witnessed an increasing interest in simpler, yet challenging, toy models where the capability of quantum simulators can be put at test against the state of the art analytical and numerical techniques.
The physics of confinement canonically belongs to high-energy physics in $3+1$ dimensions:
Quarks cannot exist in isolation and strong interaction confines them into composite particles, such as mesons and baryons.
However, confined excitations can also be realised in simple one-dimensional condensed matter settings \cite{PhysRevD.18.1259}, which are within reach of current experimental venues \cite{Martinez2016,Tan2021}. Admittedly, these realizations crudely oversimplify the complicate processes taking place in hadron colliders, but nevertheless capture many of their salient features. 
An ideal testbed for the physics of confinement is the one dimensional Ising spin chain in a tilted field \cite{DELFINO1996469,Rutkevich2008,Fonseca2003,kormos2017real}. In the presence of a pure transverse field, the Ising model is exactly solvable and equivalent to non interacting fermions. Of particular interest is the $\mathbb{Z}_2-$spontaneously broken phase, featuring two degenerate ground states of opposite magnetization. In this regime, topological excitations have the form of domain walls (or kinks) interpolating between the two vacua.\\
The nature of excitations dramatically changes in the presence of an additional longitudinal field that explicitly breaks the aforementioned symmetry by inducing a Zeeman-like gap in energy between the two lowest levels, which split in a lower and higher energy states, dubbed true vacuum and false vacuum respectively. This means that the spin chain pays energy every time it visits the false vacuum and that domain walls experience a potential that grows linearly in their separation. As a result, the kinks (playing the analogous role of quarks) cannot be pulled infinitely far apart without breaking energy conservation and get confined in composite excitations, which are identified as mesons.

The appealing simplicity of the model is self-evident and, together with other comparably simple spin chains, has been used to probe several aspects of confinement physics.
Already the pioneering paper \cite{kormos2017real} shows how confinement strongly suppresses the spreading of correlation after a quantum quench.
Shortly after, connections with quantum scarring \cite{PhysRevB.99.195108,PhysRevLett.122.130603} and lattice gauge theories \cite{RevModPhys.51.659,PhysRevB.102.041118,PhysRevLett.124.120503,ch2019confinement,pp-20} have been unveiled. In general, confinement has been understood to have striking consequence on transport \cite{PhysRevB.102.041118,PhysRevB.99.180302,bastianello2021fragmentation} and features anomalously slow thermalization due to a strong suppression of meson creation/annihilation mediated by the Schwinger mechanism \cite{PhysRev.82.664,Sinha_2021,lagnese2021false,Javier_Valencia_Tortora_2020,rigobello2021entanglement,PhysRevB.102.014308,mrw-17,cr-19,vovrosh2021}. Notably, recent works closely mimic the large hadron physics by studying scattering events of mesonic particles \cite{surace2021scattering,karpov2020spatiotemporal,milsted2021collisions,vovrosh2021confinement}.
The stepping stone for these investigations is the possibility of efficiently simulate physics in one dimension, thanks to tensor network techniques \cite{SCHOLLWOCK201196}. Numerical investigations backed up qualitative or semi quantitative analytical results, contributing to clarify the global picture. In contrast, the strongly interacting nature of these models makes fully analytical treatments scarce, especially in nonequilibrium settings.

This paper aims to analytically quantify the information spreading in confined spin chains after a quantum quench by focusing on the dynamics of entanglement.
In the framework of confinement, analytical predictions for the entanglement in the Ising chain close to the critical point have been obtained in Ref.~\cite{PhysRevLett.124.230601} in the field theory limit, but the far more common gapped scenario remains unexplored and motivates our investigation.  In the following, we first build on the standard quasiparticle picture \cite{PhysRevLett.96.136801,Calabrese2005,c-20,Alba7947} by developing a semiclassical approach able to capture the behavior of entanglement. Such semiclassical method is valid in the regime of large quantum numbers, which is achieved with a small confining force. Away from this regime, quantum effects lead to clear interference patterns in the entanglement entropy, as already observed in Ref.~\cite{kormos2017real}. In order to capture this features, we develop a fully-quantum analytical treatment that reproduces the numerical data with great accuracy. 
\begin{figure}[t]
\centering
\includegraphics[width=0.9\columnwidth]{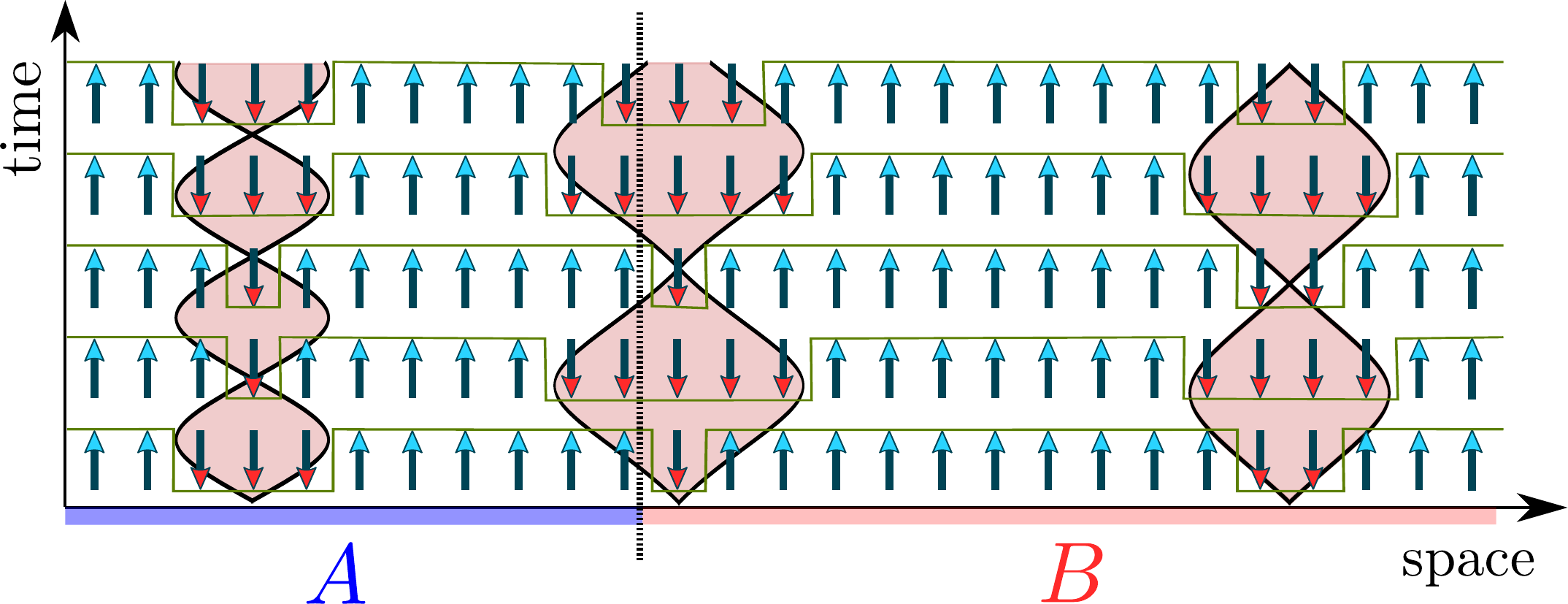}
	\caption{Illustration of the quasiparticle picture in the confined model: Small quenches mostly excite pairs of domain walls with zero total momentum, which are then confined into mesons with zero velocity. In the dilute regime, mesons behave as stationary extended particles with semiclassical trajectory $d_t(k)$ in Eq.~\eqref{eq_cl_d}. The entanglement entropy measures the shared information among the two subsystems $A\cup B$ mediated by the mesons sat at the boundaries between the two regions.}
	\label{fig_cartoon}
\end{figure}
\section{Model and quench protocol}\label{model-quench}
As a prototypical model of confined spin chain, we shall focus on the Ising chain in a tilted magnetic field with Hamiltonian
\be\label{eq_Ising}
\hat{H}=-J\sum_{j\in \mathbb{Z}}\left(\hat\sigma_j^z\hat\sigma_{j+1}^z +\hpe\hat\sigma_j^x+\hpa \hat\sigma_j^z\right),
\ee
where $\hpa$ (resp.~$\hpe$) are longitudinal (resp.~transverse) components of the magnetic field and $J$ is the overall energy scale that we set hereafter to one. We denote with $\hat\sigma_j^\alpha$ ($\alpha=x,y,z$) the standard Pauli operators acting on site $j$. For this model, we consider the following quench protocol. At times $t\leq 0$, we prepare the Ising chain in the ferromagnetic ground state by setting the transverse field to $0\leq\bar{h}_\perp<1$ at zero longitudinal field $\hpa =0$. For $t>0$, the system is then brought out of equilibrium by the sudden variation of the transverse field $\bar{h}_\perp\to\hpe$ and by turning on the longitudinal field $\hpa\ne 0$.\\
For $\hpa=0$, the model is equivalent to non interacting spinless fermions
and it allows for a well-known exact solution, see Appendix \ref{Ssec_ising_FF} for a short summary.

In particular, the Hamiltonian \eqref{eq_Ising} at $\hpa=0$ is diagonalized in terms of fermionic modes $\{\hat\gamma(k),\hat\gamma^\dagger(q)\}=\delta(k-q)$ obeying the dispersion law $\epsilon(k)=2\sqrt{(\hpe-\cos k)^2+\sin^2 k}$.
These excitations are  semiclassically interpreted as ballistic particles with velocity $v(k)=\partial_k \epsilon(k)$. 
Deep in the ferromagnetic phase $\hpe \ll 1$, the model shows two degenerate ground states (${\rm GS}_\uparrow$ and ${\rm GS}_\downarrow$, respectively) with opposite magnetization $\langle\hat\sigma^z\rangle=\pm 1$ and the excitations can be identified as sharp domain walls interpolating between the two ground states. 
At finite transverse field $\hpe<1$, the ground-state magnetization is properly renormalized $\langle\hat\sigma^z\rangle=\pm \bar{\sigma}$, with $\bar{\sigma}=(1-\hpe^2)^{1/8}$ and the fermions are still identified as dressed domain walls.\\
We further consider the confined model in Eq.~\eqref{eq_Ising} in the limit of small quenches $|\hpe-\bar{h}_\perp|,|\hpa|\ll1$, where the excitations of the gas are made of dilute mesons with zero total momentum, as depicted in Fig.~\ref{fig_cartoon}.
As matter of convention, we choose to describe low-energy excitations on the ground state with positive magnetization ($\ket{{\rm GS}_\uparrow}\equiv\ket{0}$). Indeed, the ${\rm GS}_\downarrow$ sector has the same description after replacing $\hpa\to -\hpa$. To this end, we consider the matrix elements of $\hat{H}$ on the eigenbasis of the pure transverse part, which can be labeled with asymptotic states $|\{k_i\}_{i=1}^n\rangle$ and we focus on the form factors of the longitudinal Pauli operator $\hat\sigma^z_0$, see Appendix \ref{Ssec_ising_FF} for their explicit expression.
These matrix elements can be divided in two classes: those inducing interactions within a sector with a fixed number of fermions and those violating fermion-number conservation.

A naive analysis of the Hamiltonian \eqref{eq_Ising} would put the two terms on equal footing, but effects spoiling the number conservation have been shown to be exponentially suppressed in the weak longitudinal field \cite{PhysRevB.102.041118}. Therefore, once the fermionic excitations are generated, they behave as stable particles. Nevertheless, these modes experience a non trivial interaction for $\hpa\ne 0$ due to the fermion-conserving part of the perturbation.
Such interactions are mainly of the form of a kink-antikink pair that interact through a linear potential $2\hpa \bar{\sigma}d$, where $d$ is their relative distance, plus negligible short range terms, see Appendix \ref{Ssec_ising_FF}.\
Following Ref. \cite{kormos2017real}, one can perform a quantum quench in the transverse field to create an initial distribution of fermionic excitations \cite{PhysRevLett.106.227203} and later evolve the latter in the presence of confinement. In this respect, the longitudinal field might appear as a mere spectator during the preparation of the intial state as its effect enters only at a second stage.
However, its sudden activation gives non trivial contributions to the early stage dynamics. 

Let $|\psi_0\rangle$ be the initial state, which is let to evolve with Hamiltonian \eqref{eq_Ising} as $|\psi_t\rangle= e^{-\I t \hat{H}}|\psi_0\rangle$. We split $\hat{H}$ into two parts $\hat{H}=\hat{H}_\text{diag}+\hat{H}_\text{off-diag}$, where $\hat{H}_\text{diag}$ is the projection of $\hat{H}$ onto the number conserving sector. Despite the fact that $\hat{H}_\text{off-diag}$ can be neglected in the late time dynamics, it creates excitations within a finite time scale after the quench. For this reason, it is useful to write $|\psi_t\rangle = e^{-\I t \hat{H}_\text{diag}}{\rm T}  \exp\left[-\I\int_0^t \dd \tau \, \hat{H}^{(I)}_{\text{off-diag}}(\tau)\right]|\psi_0\rangle$ where $\hat{H}^{(I)}_{\text{off-diag}}$ is the off-diagonal term of the Hamiltonian in the interaction picture with respect to the diagonal part.
Under the assumption of weak longitudinal field $\hpa\ll1$, one can argue on a separation of time scales and approximate the state as $|\psi_t\rangle \simeq e^{-\I t\hat{ H}_\text{diag}}|\tilde{\psi}_0\rangle$ by defining the renormalized initial state 
\be
|\tilde{\psi}_0\rangle\simeq {\rm T}\exp\left[-\I\int_0^{t_{\hpe}} \dd \tau \, \hat{H}^{(I)}_{\text{off-diag}}(\tau)\right]|\psi_0\rangle\, ,
\ee
with $t_{\hpe}$ a large time scale compared with the transverse Ising energy scale $t_{\hpe}\epsilon(k)\gg 1$, but still smaller when compared with the longitudinal field $t_{\hpe} \hpa\ll 1$.
In Appendix \ref{Ssec_state_preparation} we substantiate on this idea and compute the renormalized initial state $\ket{\tilde\psi_0}$ in the sought limit of small quenches using form factor perturbation theory \cite{DELFINO1996469,delfino-14,dv-14} and obtaining a simple squeezed state
\be\label{eq_squeezed}
|\tilde{\psi}_0\rangle\propto\exp\left[\int_0^\pi \dd k\,  \mathcal{K}(k) \hat\gamma^\dagger(k) \hat\gamma^\dagger(-k)\right] |0\rangle\, .
\ee
Above, mode operators act on the post-quench vacuum $\hat\gamma(k)|0\rangle=0$. The wave function $\mathcal{K}$ can be written as
\be\label{WF}
\mathcal{K}(k)=-\I\tan(\theta_k^{\hpe}-\theta_k^{\bar{h}_\perp})-\I \hpa\bar{\sigma}v(k)/\epsilon^2(k)\, ,
\ee where $\theta_k^{\hpe}$ is the Bogoliubov angle that diagonalizes the transverse part of the Hamiltonian, see Appendix \ref{Ssec_state_preparation} for the detailed derivation. Here, the energy $\epsilon(k)$ and velocity $v(k)$ are computed with respect to the post-quench transverse field $\hpe$.
The squeezed form represents an incoherent superposition of pairs of particle with opposite momenta and density $n_k=|\mathcal{K}(k)|^2(1+|\mathcal{K}(k)|^2)^{-1}$. The limit of small quenches is then attained for $|\mathcal{K}(k)|^2\ll 1$.
\section{Entanglement dynamics}\label{sec:entanglement}

Entanglement quantifies the amount of shared quantum information in a bipartition $A\cup B$. The most common entanglement measure is the von Neumann entropy $S_A=-\text{Tr}\hat\rho_A\log\hat\rho$ of the reduced density matrix $\hat\rho_A=\text{Tr}_B |\psi_t\rangle\langle \psi_t|$, but the R\'enyi entropies $S^{(n)}_A=-\frac{1}{1-n}\log(\text{Tr}\hat\rho_A^n)$ recently became experimentally accessible \cite{Islam2015,M.2016,PhysRevLett.120.050406,Tiff2019,PhysRevLett.125.200501,Neven2021,vitale2021symmetryresolved}.

The quasiparticle picture (QP) \cite{PhysRevLett.96.136801,Calabrese2005,Alba7947} judiciously supplements few quantum inputs with semiclassical arguments, providing the leading order extensive part of the entanglement entropy. Within this framework, local pairs of entangled excitations with opposite momenta are seen as an initial source for the entanglement and, during the dynamics, the propagating pairs that are shared by the two subsystems become responsible for the entanglement growth. Notably, the QP picture can be used beyond the pair structure of the post-quench state \cite{Bertini2018,10.21468/SciPostPhys.5.4.033,10.21468/SciPostPhys.8.3.045}.
In the context of confinement, these ideas give a quick qualitative grasp on the expected behavior of entanglement. 
In particular, sooner or later, the confining force bends the short time ballistic trajectories $d_t(k)\simeq 2 v(k) t$ and consequently mitigates the entanglement growth, see Fig.~\ref{fig_cartoon} for an illustration.
Focusing on a single pair created at momenta $(-k,k)$ and overlapping position, the classical distance is \cite{Rutkevich2008}
\be\label{eq_cl_d}
d_t(k)=(\hpa \bar{\sigma})^{-1}[\epsilon(k)-\epsilon(k-2\hpa \bar{\sigma}  (t\text{ mod }k/(\hpa \bar{\sigma})))]\, .
\ee
In the limit of dilute mesons, each fermionic pair remains well-separated and thus independent from others during the time evolution. Precisely, this regime is reached if $d_\text{max} \int_0^\pi\frac{\dd k}{2\pi}\, n_k\ll 1$, where $d_\text{max}=4 \hpe/( |\hpa| \bar{\sigma})$ is the maximum extension of a meson.
In the case where only the longitudinal field is quenched and $\hpe$ remains constant, Eq. \eqref{WF} leads to the simple expression $\int_0^\pi\frac{\dd k}{2\pi}\, n_k=(\hpa\hpe\bar{\sigma})^2/[16(1-\hpe^2)^3]$. Hence, the bond is made explicit and it is linearly improved in the weak longitudinal field.
Conversely, at higher densities, the single-meson physics breaks down due to collisions among mesons. 
We shall focus on the dilute regime, where the standard quasiparticle prediction for the entanglement entropy can be readily generalized to the confined model by replacing the term $2t v(k)$ with $d_t(k)$ given in Eq.~\eqref{eq_cl_d}. In the case of an infinite system divided in two semi-infinite halves $A\cup B=(-\infty,0]\cup (0,+\infty)$, the von Neumann entropy is given by the formula
\be\label{eq_QP_entropy}
S_A(t)=\int_0^\pi \frac{\dd k}{2\pi} d_t(k) [-n_k\log n_k
 -(1-n_k)\log(1-n_k)]\, .
\ee
\begin{figure}[t!]
	\includegraphics[width=\columnwidth]{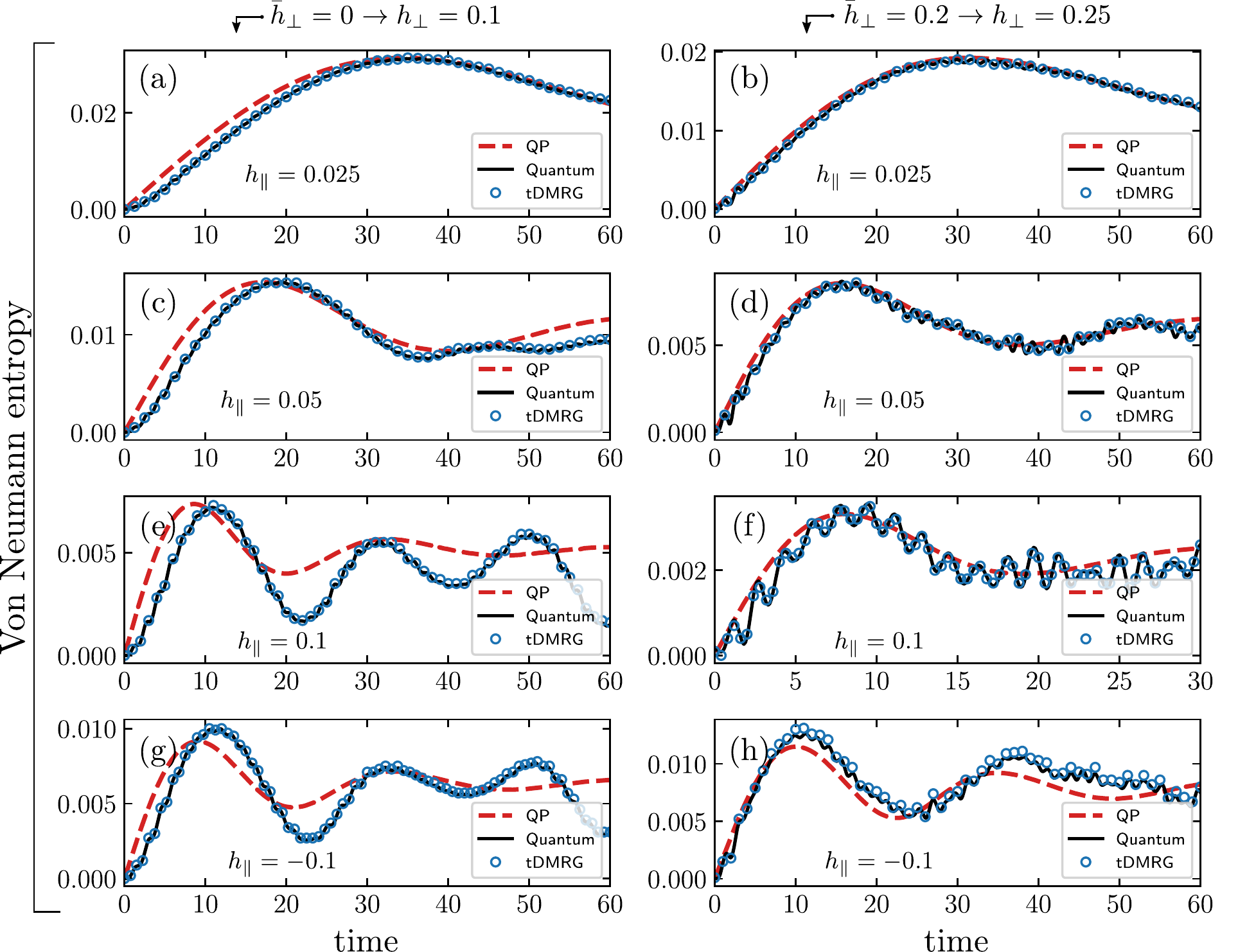}
	\caption{von Neumann entropy dynamics of the confined Ising model \eqref{eq_Ising}. We show different quenches of the transverse field on different columns and different values of $\hpa$ on different rows. In each panel, we compare the semiclassical prediction \eqref{eq_QP_entropy} ({\it dashed line}), the quantum prediction in Gaussian approximation ({\it full line}) and numerical data obtained with time-dependent DMRG simulations ({\it symbols}). Notice the validity of the approach also for quenches starting from the false vacuum (g)-(h).}
	\label{fig:results}
\end{figure}
In Fig.~\ref{fig:results} we test the semiclassical prediction in Eq.~\eqref{eq_QP_entropy} against tensor network simulations for different quench parameters, finding  a very good agreement in the regime of weak longitudinal field $\hpa\ll 1$.
Notice that the applicability of our analysis holds also for negative longitudinal fields, namely when mesons are true-vacuum excitations in a {\it false vacuum} sea, see Fig.~\ref{fig:results} (g)-(h). In this case, the linear potential changes sign and becomes repulsive, but the lattice induces Bloch oscillations preventing the domain walls to be repelled infinitely far apart (see also Ref. \cite{pomponio2021bloch}). By increasing the value of the longitudinal field, deviations from Eq.~\eqref{eq_QP_entropy} eventually undermine the approximation of isolated mesons, see Appendix \ref{app:breakdown} for further details.
However, we observe a failure of the semiclassical approximation far before this limiting case. Indeed, in Eq.~\eqref{eq_cl_d} $d_t(k)$ is treated as a continuum variable and, a consequence, Eq.~\eqref{eq_QP_entropy} is expected to hold only in the limit $d_\text{max}\gg 1$. Needless to say, a simple theory outclassing this semiclassical treatment and including quantum effects is highly desirable.
\subsection{Quantum effects beyond the quasiparticle picture}\label{sec:Quantum}
Computing the entanglement of many-body systems is an extremely hard task with very few notable exceptions where analytical derivations are possible, including e.g. critical models \cite{Calabrese2004,Calabrese2009} and non interacting systems \cite{Casini2009}. While in Ref.~\cite{PhysRevLett.124.230601}, conformal invariance has been exploited for similar purposes,
we rather borrow methods usually applied to non interacting models.
Let us consider a system of spinless fermionic operators $\hat{c}_j$, $\hat{c}^\dagger_j$, which will eventually be chosen as the Jordan-Wigner fermions of the Ising chain \eqref{eq_Ising}, see Appendix \ref{Ssec_EE}.
The distinctive feature of a non interacting system is the applicability of Wick's theorem, which holds
for any Gaussian state. Gaussian states are fully characterized by their quadratic correlations
\be\label{eq_C}
C=\begin{pmatrix} \langle \hat{c}_j \hat{c}^\dagger_{j'}\rangle && \langle \hat{c}_j \hat{c}_{j'}\rangle \\
\langle \hat{c}^\dagger_{j} \hat{c}^\dagger_{j'}\rangle && \langle \hat{c}^\dagger_j \hat{c}_{j'}\rangle\end{pmatrix}
\ee
From the knowledge of $C$, one can easily construct the correlation matrix $C_A$ restricted to the subsystem $A$ and evaluate the entanglement entropy as $S_A=-\text{Tr}[C_A\log C_A]$ \cite{Casini2009}. Notice that the trace is over the lattice sites of $A$, thus the dimension of the vector space scales linearly with the size of $A$, rather than exponentially as the dimension of the Hilbert space describing the subsystem. Therefore, the entanglement entropy is obtained in a matter of seconds on a laptop.\\
Notably, the initial state in Eq.~\eqref{eq_squeezed} is Gaussian. We now argue that the dilute regime is compatible with a Gaussian-preserving dynamics. Indeed, since Eq.~\eqref{eq_squeezed} describes an incoherent superposition of paired fermions that evolve independently, it is expected to capture dilute mesons provided the single-meson dynamics is properly described by  a non trivial  time-dependent wave function $\mathcal{K}_t(k)$.
The wave function $\mathcal{K}_t(k)$ is obtained from the Schr\"odinger equation $\I\partial_t |\psi_t\rangle= \hat{H}_\text{off-diag}|\psi_t\rangle$ by projecting onto the two-particle subspace. The details of this derivation are lengthy but straightforward and can be found in Appendix \ref{Ssec_ising_FF}. The resulting equation is conveniently written in real space as
\be\label{eq_twobody}
\I\partial_t W_t(j)=2\hpa \bar{\sigma} |j|W_t(j)+\sum_{\ell\in\mathbb{Z}} [T_{j-\ell}-T_{-j-\ell}] W_t(\ell)\, ,
\ee
where $W_t(j)$ is an antisymmetric wave function related to ${\cal K}_t(k)$ as $W_t(j)=\int \frac{\dd k}{2\pi}e^{-\I k j} \mathcal{K}_t(k)$ and $T_j=\int \frac{\dd k}{2\pi}  e^{-\I k j}2\epsilon(k)$ is the kinetic energy term.
Notice that in the regime of weak transverse field $\epsilon(k)\simeq 2- 2\hpe \cos k$, $T_j$ reduces to a first neighbor hopping and the Schr\"odinger equation \eqref{eq_twobody} can be exactly diagonalized \cite{Rutkevich2008}. For finite transverse fields, the one-body problem can be easily tackled numerically.
For completeness, it should be mentioned that additional small short-range interactions are present in Eq.~\eqref{eq_twobody}, see Appendix~\ref{Ssec_ising_FF} for details.
In Fig.~\ref{fig:results} we compare tensor network simulations against the quantum prediction obtained in Gaussian approximation, finding an excellent agreement. Technical details on the numerical calculations and the comparison with tensor network simulations are reported in Appendix \ref{Ssec_EE}.

\begin{figure}[t!]
\centering
\includegraphics[width=\columnwidth]{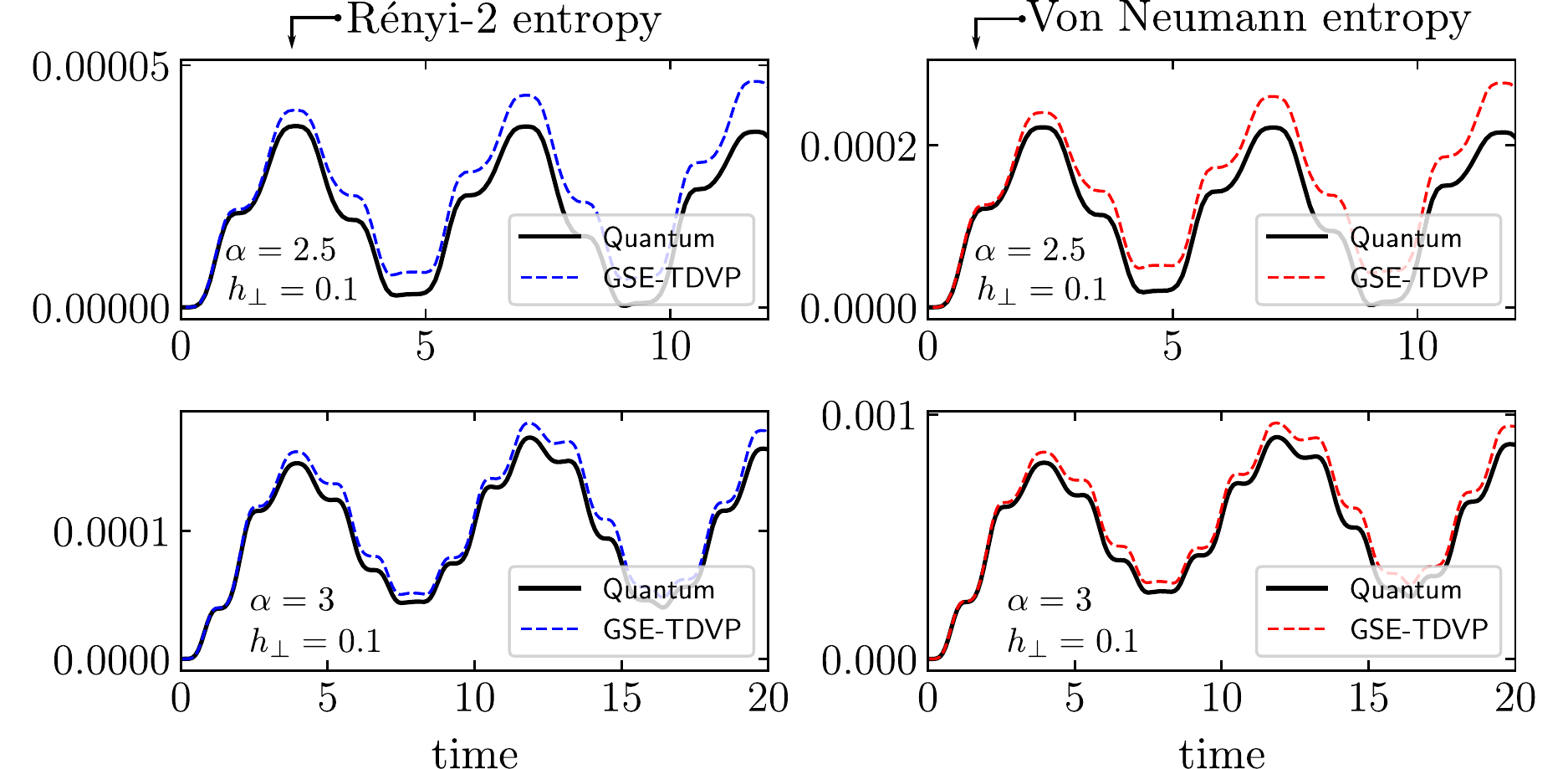}
\caption{(Left) R\'enyi-2 and (right) von Neumann entropy dynamics for the long-range Ising model \eqref{eq_longH} with different exponent $\alpha$ on different rows. The system is initially prepared in a ferromagnetic state with all spins up. The quantum prediction in Gaussian approximation ({\it full line}) is compared with numerical data ({\it dashed line}) obtained with global subspace expansion time-dependant variational principle \cite{GSE-TDVP}.}\label{fig:long_range}
\end{figure}

\section{Confined spin chains in experiments}\label{sec:long-range}
Recent experimental platforms of trapped ions show clear signature of confined physics \cite{Tan2021}. In this setups, the confinement is induced by the long-range interactions of the Ising Hamiltonian \cite{PhysRevLett.122.150601}
\be\label{eq_longH}
\hat{H}=-\sum_{i<j}\frac{1}{|i-j|^\alpha}\hat\sigma^z_i\hat\sigma_j^z-\hpe \sum_i \hat\sigma_i^x\, .
\ee
For small transverse field $\hpe$, the fundamental excitations of the model \eqref{eq_longH} are domain walls, analogously to the short-range Ising chain in Eq.~\eqref{eq_Ising} we previously discussed. In view of Eq. \eqref{eq_longH}, a pair of domain walls experiences an attractive potential that for $1<\alpha\le2$ indefinitely grows as the distance between the particles increases. In this way, a confining mechanisms is induced.
For $\alpha> 2$, the intra-kink interaction is bounded, but it still supports deep bound states with meson-like dynamics for $\alpha$ not too large.
The methods discussed previously for the Hamiltonian \eqref{eq_Ising} can be easily generalized to the long-range Ising model \eqref{eq_longH}. 
In Fig.~\ref{fig:long_range}, we initialize the system in a product state of all spins up and we let it evolve with the Hamiltonian \eqref{eq_longH} for small $\hpe$. 
Due to its experimental relevance \cite{Islam2015,M.2016,PhysRevLett.120.050406,Tiff2019,PhysRevLett.125.200501,Neven2021,vitale2021symmetryresolved}, we also show the results for the R\'enyi $S^{(2)}_A$ which, analogously to the Von Neumann entropy, is accessible from correlation functions \eqref{eq_C} (see Appendix \ref{Ssec_long_range}).
In contrast with the longitudinal case, the long-range Hamiltonian \eqref{eq_longH} induces meson-meson and boundary-meson interactions beyond short range terms, harming the approximation of dilute, thus non interacting, mesons. As a consequence of interactions, mesons become mobile causing an additional linear drift of the entanglement over the single-meson physics result.
As expected, the drift is more evident as $\alpha$ is smaller, see Fig.~\ref{fig:long_range}.
Further details on the state preparation and on the breakdown our approximations are left to Appendices \ref{Ssec_long_range} and \ref{app:breakdown}.

\section{Conclusions}
We provide an analytical study of the entanglement dynamics in confined spin chains based on a minimal set of ingredients. Precisely: (i) mesons are incoherently excited and non interacting, hence squeezed states remain a good approximation also at late times and (ii) Gaussian correlators are evolved according to the dynamics projected in the two-fermion sector.
Although our results have been explicitly worked out only for Ising spin chains with short and long range interactions, it is clear that our methods 
apply in general, as e.g. in confining spin ladders \cite{lsmc-21b,lskc-20}, Potts models \cite{lt-15,dg-08} and tricritical Ising \cite{lencses2021confinement}.
Natural future directions address the breakdown of the single-meson approximation and the effect of moving \emph{---}and hence scattering\emph{---} mesons on the entanglement dynamics. Particularly appealing is the semiclassical limit, where entangled particles move along classical trajectories in the spirit of \cite{Bertini2018,10.21468/SciPostPhys.5.4.033,10.21468/SciPostPhys.8.3.045,PhysRevLett.119.100603}. Quantifying how the mesons scatter would eventually allow us to understand whether 
these confining models thermalize.

\section*{Acknowledgements}
AB acknowledges support from the Deutsche Forschungsgemeinschaft (DFG, German Research Foundation) under Germany's Excellence Strategy -- EXC–2111–390814868. PC and SS acknowledge support from ERC under Consolidator Grant No. 771536 (NEMO). The authors acknowledge Mario Collura for useful discussions and insights on the numerical simulations. SS is grateful to Lorenzo Piroli for discussions and to Emanuele Tirrito for help with TDVP simulations. AB is grateful to Alessio Lerose, Michael Knap, Stefan Birnkammer, Johannes Knolle, Joseph Vovrosh and Hongzheng Zhao for several discussions and their insight on related work.
Numerical simulations have been performed with the open-source library iTensor \cite{itensor}.

\appendix

\section{The Ising chain in a weak longitudinal field: the form factor approach}
\label{Ssec_ising_FF}

We consider the pure transverse field Ising chain
\be\label{Ising}
\hat{H}_\text{trans}=- \sum_{j\in\mathbb{Z}} \left(\hat\sigma^z_{j+1}\hat\sigma_j +h_\perp \hat\sigma^x_j\right)\,,
\ee
which can be exactly diagonalized in terms of non interacting spinless fermions as we now briefly review. For simplicity, we shall describe the Ising model on a lattice of $N$ sites, assuming periodic boundary conditions of the chain and eventually we take the thermodynamic limit where $N\to \infty$. First, we introduce fermionic creation and annihilation operators $\hat{c}^\dagger_j$, $\hat{c}_j$ by means of a Jordan-Wigner transformation of the ladder Pauli operators as
\be
\frac{\hat\sigma_j^x+\I\hat\sigma_j^y}{2}=\exp\big(\I\pi \sum_{i<j} \hat{c}^\dagger_{i}\hat{c}_{i}\big)\hat{c}^\dagger_j\,. 
\ee
With this transformation, it is possible to divide the Ising Hamiltonian \eqref{Ising} in two terms $\hat{H}_\text{trans}^\pm$ according with the parity operator $\hat{P}=\prod_j \hat\sigma_j^x$ (notice that $\hat{P}$ has eigenvalues $P=\pm 1$ and it commutes with the transverse Hamiltonian)
\be
\hat{H}_\text{trans}=\frac{1}{2}(1+\hat{P}) \hat{H}_\text{trans}^++\frac{1}{2}(1-\hat{P})  \hat{H}_\text{trans}^-\, .
\ee
In this form, the Ising Hamiltonian becomes quadratic in the fermionic operators and translationally  invariant in each of the two parity sectors. Notice that the terms $\hat{H}_\text{trans}^\pm$ differ only for the boundary terms: The parity sector $P=+1$ ($P=-1$) is characterized by antiperiodic (periodic) boundary conditions for the fermionic operators and it induces a quantization of momenta in half-integer (integer) multiples of $2\pi/N$. Nevertheless, these boundary effects become negligible in the thermodynamic limit and will be thus omitted in what follows. Hence, in both parity sectors the Hamiltonian can be diagonalized in the momentum space as
\be
\hat{H}_\text{trans}^{\pm}=\int \dd k \ \epsilon(k) \hat\gamma_\pm^\dagger(k)\hat\gamma_\pm(k)+\ \text{const.}
\ee
with single-particle energy 
\be
\epsilon(k)=2\sqrt{(\hpe-\cos(k))^2+\sin^2 k}.
\ee
The new set of operators  $\hat\gamma_\pm$ is defined by the Bogoliubov rotation 
\be\label{S_eq_bog_rot}
\begin{pmatrix}\hat{c}_j \\[4pt] \hat{c}^\dagger_j \end{pmatrix}=\int \frac{\dd k}{\sqrt{2\pi}} \; U_{\theta_k}\begin{pmatrix} \hat\gamma_\pm(k) \\[4pt] \hat\gamma_\pm^\dagger(-k)\end{pmatrix}\ee
where 
\be
U_{\theta_k}=\begin{pmatrix} \cos\theta_k && \I\sin\theta_k \\ \I\sin\theta_k && \cos\theta_k\end{pmatrix}
\ee
and $\theta_k$ is the Bogoliubov angle
\be\label{Seq_Bog_angle}
\theta_k=-\frac{1}{2\I}\log\left(\frac{h_\perp-e^{\I k}}{(\cos k-h_\perp)^2+\sin^2k}\right)\, .
\ee
With our conventions for the Fourier transform, the new modes obey standard anticommutation relations $\{\hat\gamma_\pm(k),\hat\gamma^\dagger_\pm(q)\}=\delta(k-q)$ for any value of $\theta_k$. In each of the two parity sectors, the Hilbert space is understood as a Fock space built on the vacua $|0_\pm\rangle$ that are annihilated by the respective mode operators $\hat\gamma_\pm(k)|0_\pm\rangle=0$.
For $h_\perp<1$, the model undergoes spontaneous symmetry breaking in the thermodynamic limit. This can be easily understood in the classical limit $h_\perp\to 0$, where there are two degenerate ground states $|{\rm GS}_\uparrow\rangle=|...\uparrow\uparrow\uparrow...\rangle$ and $|{\rm GS}_\downarrow\rangle=|...\downarrow\downarrow\downarrow...\rangle$. Through linear combinations of these two vacua, one forms $P$-symmetric and antisymmetric pairs, which are then identified with the vacua $|0_\pm\rangle=(|{\rm GS}_\uparrow\rangle \pm |{\rm GS}_\downarrow\rangle)/\sqrt{2}$ with opposite value of magnetization $\bar{\sigma}=\pm1$. At finite $h_\perp$, the degenerate ground states are not simple product states any longer and the longitudinal magnetization gets renormalized to the value $\bar{\sigma}\equiv\langle {\rm GS}_\uparrow|\sum_{j=1}^N \hat\sigma^z_j/N|{\rm GS}_\uparrow\rangle=-\langle {\rm GS}_\downarrow|\sum_{j=1}^N\hat\sigma^z_j/N|{\rm GS}_\downarrow\rangle=(1-h_\perp^2)^{1/8}$ \cite{Rutkevich2008}.

\subsection{The form factor approach to the longitudinal field}
We now wish to consider the effect of the longitudinal field $\hpa$ in Eq.~\eqref{eq_Ising} on the low-energy excitations above one of the degenerate vacua. For the sake of concreteness, we choose the ground state with positive magnetization $|{\rm GS}_\uparrow\rangle$. As a first step, one builds the proper excitation basis on the Fock space of the transverse part, by considering the symmetric combination of the Fock spaces with opposite parity. More precisely, let $|(\{k_i\}_i^N)_\pm\rangle$ be a multiparticle state in the $P=\pm 1$ sectors, then one defines $|\{k_i\}_{i=1}^N\rangle= \frac{1}{\sqrt{2}}(|(\{k_i\}_i^N)_+\rangle+|(\{k_i\}_i^N)_-\rangle)$. In this notation, the state with no excitations is the desired ground state, i.e., $|0\rangle\equiv |{\rm GS}_\uparrow\rangle$. Notice that the analysis at finite system size $N$ is more involved due to the different quantization conditions of the even and odd parity sectors. The action of the confined Ising Hamiltonian in Eq.~\eqref{eq_Ising} on this basis is
\begin{widetext}
\be
\hat{H}|\{k_i\}_{i=1}^N\rangle= \left(\sum_{i=1}^N \epsilon(k_i)\right)|\{k_i\}_{i=1}^N\rangle+\hpa\sum_{M=1}^\infty \frac{1}{M!}\int_{-\pi}^\pi \frac{\dd^M q}{(2\pi)^M}\, 2\pi\delta\left(\sum_{j=1}^M q_j-\sum_{i=1}^N k_i\right) |\{q_j\}_{j=1}^M\rangle \langle\{q_j\}_{j=1}^M|\hat\sigma^z_0|\{k_i\}_{i=1}^N\rangle\, .
\ee
\end{widetext}
In this expression, the form factors $\langle\{q_j\}_{j=1}^M|\hat\sigma^z_0|\{k_i\}_{i=1}^N\rangle$ that appear were previously determined in Ref.~\cite{Rutkevich2008}. By a repeated use of Wick theorem, one can be express the latter in terms of the following two particle form factors: 
\begin{subequations}
\be
\langle k_1 k_2|\hat\sigma^z_0|0\rangle= \bar{\sigma} F(k_1,k_2|);
\ee
\be \langle 0|\hat\sigma^z_0|k_1, k_2\rangle= \bar{\sigma} F(|k_1, k_2); 
\ee
\be \langle k_1|\sigma_0^z|k_2\rangle=\bar{\sigma}F(k_1|k_2),
\ee
\end{subequations}
 where
\be\label{S_eq_ff}\begin{split}
&F(k_1,k_2|)=-F^*(|k_1,k_2)\\[5pt]
&\quad=\frac{1}{1-\exp[\I(k_1+k_2)]}\frac{\epsilon(k_1)-\epsilon(k_2)}{\sqrt{\epsilon(k_1)\epsilon(k_2)}}\,
\end{split}\ee
and
\be
F(k_1|k_2)=\frac{1}{1-\exp[\I(k_1-k_2)]}\frac{\epsilon(k_1)+\epsilon(k_2)}{\sqrt{\epsilon(k_1)\epsilon(k_2)}}\, .
\ee
Notice that $F(|k_1,k_2)$ is always finite while $F(k_1|k_2)$ develops a kinematic singularity for $k_1=k_2$ that has to be properly regularized by a symmetric shift in the imaginary axis
\be\label{eq_S9}\begin{split}
&\frac{2}{1-\exp[\I(k_1-k_2)]}\to \left[\frac{2}{1-\exp[\I(k_1-k_2)]}\right]_{\text{reg}} \\[5pt]
&=\frac{1}{1-\exp[\I(k_1-k_2)]+\I 0^+}+\frac{1}{1-\exp[\I(k_1-k_2)]-\I 0^+}\, .
\end{split}\ee

\subsection{The dynamics in the two particle sector} 
The singular part of $F(k_1|k_2)$ is responsible for the long-range confinement, as it is clarified by the dynamics in the two-fermion sector that we now analyze. Hence, we now focus on the matrix elements of the form $\langle q_1,q_2|\hat{H}|k_1,k_2\rangle$.
In particular, the matrix element of the longitudinal Pauli operators reads
\be\begin{split}
&\bar{\sigma}^{-1}\langle q_1 q_2|\hat\sigma_0^z|k_1,k_2\rangle =F(q_1,q_2|)F(|k_1,k_2)\\[3pt]
&\qquad +F(q_1|k_2)F(q_2|k_1) -F(q_1|k_1)F(q_2|k_2)\, .
\end{split}\ee
We further specialize to the zero-momentum sector where one sets $(k_1,k_2)=(k,-k)$ and similarly $(q_1,q_2)=(q,-q)$, due to momentum conservation. In this case, we obtain $F(|k,-k)=\I v(k)/\epsilon(k)$, while the singularity in Eq. \eqref{eq_S9} requires a more careful treatment. The calculation are lengthy but straightforward, hence we report below the final result only (see Ref.~\cite{Rutkevich2008} for further details).
We define the two-body interaction in the momentum space as $V(q,k)\equiv \bar{\sigma}^{-1}\langle q,-q|\hat\sigma_0^z|k,-k\rangle$ and we divide the regular part from the singular part as
\be\begin{split}
&V(q,k)=V^\text{reg}(q,k)+\\[4pt]
&\quad\left[\frac{2}{1-\exp[\I(q+k)]}\right]_\text{reg}\left[\frac{2}{1-\exp[\I(-q-k)]}\right]_\text{reg}\\[4pt]
&\quad-\left[\frac{2}{1-\exp[\I(q-k)]}\right]_\text{reg}\left[\frac{2}{1-\exp[\I(-q+k)]}\right]_\text{reg}\, .
\end{split}\ee
The double pole singularity is responsible for the linear confinement, while $V^\text{reg}(q,k)$ is not singular and gives rise to a short-range interaction. Next, we introduce the wave function ${\cal K}(k)$ and we write the two-body state as
\be
|\psi_{2-\text{fermions}}\rangle=\frac{1}{2}\int_{-\pi}^\pi \dd k\,  \mathcal{K}(k) |k,-k\rangle\, ,
\ee
where the notation is chosen to match the expansion of the squeezed state in Eq.~\eqref{eq_squeezed}.
Furthermore, due to the long-range nature of confinement, it is convenient to move to coordinate space by defining the real-space wave function $W(j)$ as
\be
W(j)=\int \frac{\dd k}{2\pi}e^{-\I k j} \mathcal{K}(k)\, .
\ee
 Hence, projecting the Schr\"odinger equation to the two-particles state and by evaluating the action of the Hamiltonian in Eq.~\eqref{eq_Ising} on the wave function $W(j)$, one obtains
\begin{widetext}\be\label{S_eq_Sh}
\I\partial_t W(j)=2\hpa \bar{\sigma}|j|W(j)+\sum_{\ell>0}\Big[ T_{j-\ell}-T_{-j-\ell} -T_{j+\ell} +T_{-j+\ell}+\frac{\hpa\bar{\sigma}}{2} V^\text{reg}_{j,\ell} \Big] W(\ell)\, ,
\ee
\end{widetext}
with $T_j=\int_{-\pi}^\pi \frac{\dd k}{2\pi} e^{-\I k j} 2\epsilon(k)$ and 
\be
V^\text{reg}_{j,\ell}=\int \frac{\dd k \dd q}{(2\pi)^2} e^{-\I kj}e^{\I q \ell} V^\text{reg}(k,q)\, . 
\ee
Notice that $V^\text{reg}_{j,\ell}$ is a short range interaction that vanishes whenever $j$ and $\ell$ are large, regardless their relative difference. While in principle both the linear potential and the short term interactions are of order ${\cal O}(\hpa)$, the long-ranged nature of the first makes $V^\text{reg}_{j,\ell}$ negligible in most of practical cases.

\section{State preparation thorough a quantum quench}
\label{Ssec_state_preparation}

In this section, we characterize the state obtained after a quantum quench in the transverse and the longitudinal magnetic field. Let $|\psi_0\rangle$ be the initial state, set as the ground state of the model for a certain value of transverse field $\bar{h}_\perp$ at zero longitudinal coupling. To begin with, we express $\ket{\psi_0}$ in the basis of the post-quech fermions, which amounts to solve the quench in the transverse field \cite{PhysRevLett.106.227203}. From Eq.~\eqref{S_eq_bog_rot}, one can relate the pre- and post-quench modes $\hat{\bar\gamma}(k)$ and $\hat\gamma(k)$ respectively  by a simple rotation 
$\hat{\bar\gamma}(k)=\cos(\theta_k^{\hpe}-\theta_k^{\bar{h}_\perp})\hat\gamma(k)+\I\sin(\theta_k^{\hpe}-\theta_k^{\bar{h}_\perp})\hat\gamma^\dagger(-k)$.
Imposing that $\hat{\bar\gamma}(k)|\psi_0\rangle=0$, one obtains a simple equation for the post-quench modes, which has the following solution
\be\label{eq:psi_0}
|\psi_0\rangle=\frac{1}{\sqrt{\mathcal{N}}}\exp\left[\int_0^\pi \dd k\,K(k) \hat\gamma^\dagger(k)\hat\gamma^\dagger(-k)\right]|0\rangle\, ,
\ee
where $K(k)=-\I \tan(\theta_k^{\hpe}-\theta_k^{\bar{h}_\perp})$ and $\mathcal{N}$ is a normalization factor. However, as anticipated in Sec.~\ref{model-quench}, the activation of a longitudinal magnetic field has non trivial consequences on the state that we now address by computing the renormalized initial state
\be\label{eq:ren-psi_0}
|\tilde{\psi}_0\rangle\simeq {\rm T}\exp\left[-\I\int_0^{t_{\hpe}} \dd \tau \, \hat{H}^{(I)}_{\text{off-diag}}(\tau)\right]|\psi_0\rangle.
\ee
In the limit of small quench where $K(k)$ is small, the initial state \eqref{eq:psi_0} can be expanded as
\be
|\psi_0\rangle \propto |0\rangle +\int_0^\pi \dd k\, K(k)|k,-k\rangle+ \dots 
\ee
and, at leading order in $\hpa\ll 1$, one obtains from \eqref{eq:ren-psi_0}
\be\label{S_eq_psi0exp}\begin{split}
|\tilde{\psi}_0\rangle&= \frac{1}{\sqrt{\mathcal{N}}}\Big( |0\rangle+\int_0^\pi \dd k\, K(k)|k,-k\rangle\\
&\qquad -\I \int_0^{t_{\hpe}} \dd \tau \ \hat{H}^{(I)}_{\text{off-diag}}(\tau)|0\rangle+\dots\Big)\, .
\end{split}\ee
In principle, $\hat{H}^{(I)}_{\text{off-diag}}(\tau)|0\rangle$ can couple to a sector with an arbitrary even number of fermions, however in the small quench limit $\hpe\ll 1$ the most important contribution comes from the two-particle sector. This can be argued on the basis of two facts: (i) as we will see, the cross ratio of a particles pair creation is associated with the form factors $F(k_1,k_2|)$ in Eq.~\eqref{S_eq_ff}, which for small transverse field scale as $\sim \hpe$, hence couplings to the sector with $2n$ fermions are of order $\sim \hpe^n$; (ii) any process is multiplied by the inverse of the energy of the final state, which means that couplings to sectors with more fermions are further suppressed.
Motivated by this reasoning, we now wish to compute the quantity $ \int_0^{t_{\hpe}} \dd \tau \langle k_1,k_2|\hat{H}^{(I)}_{\text{off-diag}}(\tau)|0\rangle$.
As a further approximation, it is expected that for $\hpa\ll 1$ fermions are locally created on a much shorter time scale compared to the one over which the confining force shows appreciable effects. Therefore, we write the Hamiltonian in the interaction picture as $\hat{H}^{(I)}_{\text{off-diag}}(\tau)\simeq e^{\I\tau H_\text{trans}}\hat{H}_{\text{off-diag}}e^{-\I\tau \hat{H}_\text{trans}}$, neglecting the term of confinement.
Under this program, we finally obtain the expression
\be\begin{split}
&-\I\int_0^{t_{\hpe}} \dd \tau \langle k_1,k_2|\hat{H}^{(I)}_{\text{off-diag}}(\tau)|0\rangle\simeq\\
&-\I\int_0^{t_{\hpe}} \dd \tau  \hpa \bar{\sigma} e^{\I\tau( \epsilon(k_1)+\epsilon(k_2))}\left[\sum_j e^{\I(k_1+k_2)j}\right] F(k_1,k_2|)\\[4pt]
&\stackrel{\epsilon(k)t_{\hpe}\gg 1}{=}-\I2\pi \delta(k_1+k_2) \hpa \bar{\sigma}\frac{v(k)}{\epsilon^2(k)}.
\end{split}\ee
In the last passage, the large time limit is performed in the distribution sense. Plugging this result in Eq. \eqref{S_eq_psi0exp}, we obtain $|\tilde{\psi}_0\rangle\simeq \frac{1}{\sqrt{\mathcal{N}}}\left( |0\rangle+\int_0^\pi \dd k\, \mathcal{K}(k)|k,-k\rangle+\dots\right)$ with a renormalized wave function $\mathcal{K}(k)=-\I\tan(\theta_k^{\hpe}-\theta_k^{\bar{h}_\perp})-\I \hpa\bar{\sigma}v(k)/\epsilon^2(k)$ (cf with Eq.~\eqref{WF}).

\begin{figure}[t!]
\centering
\includegraphics[width=\columnwidth]{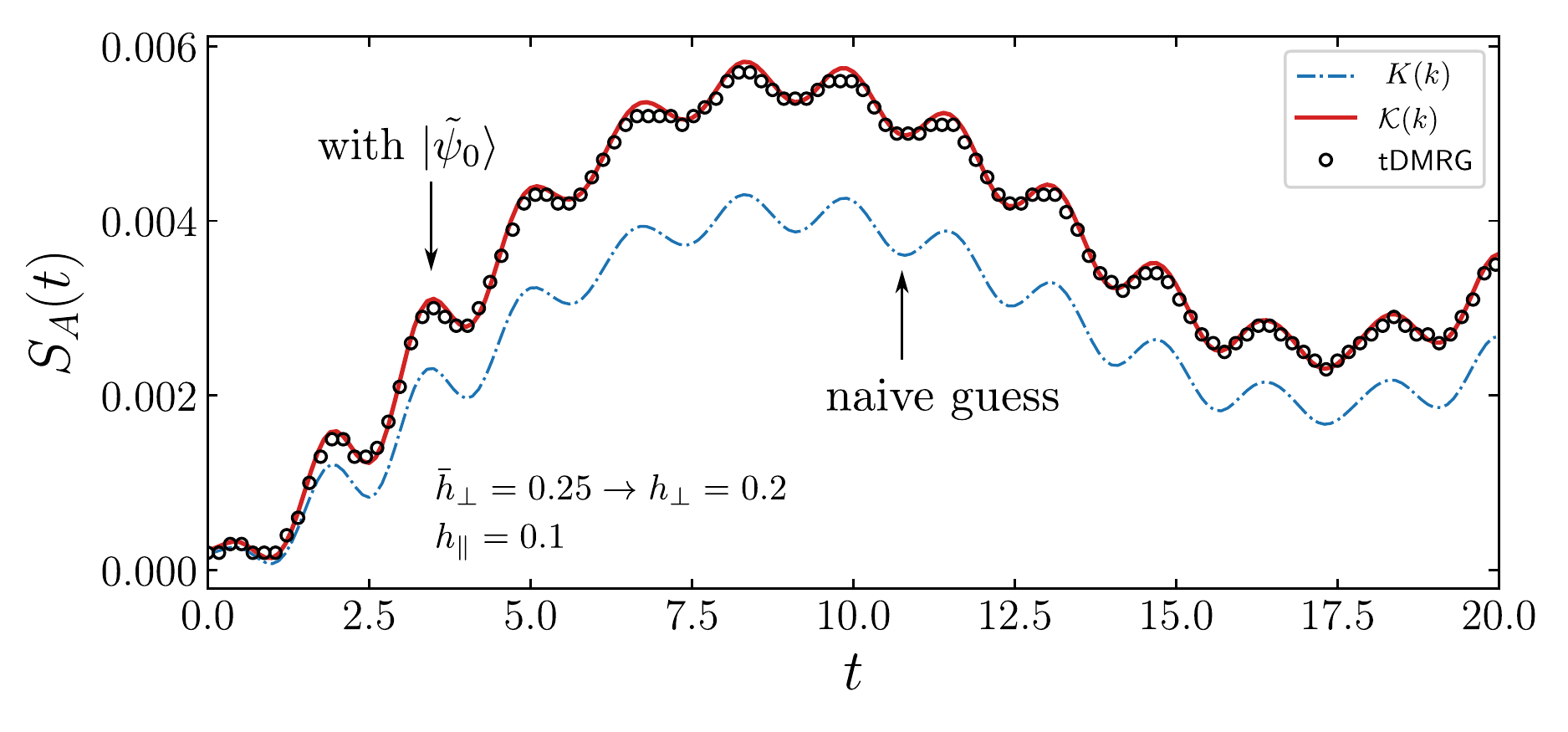}
\caption{Comparison of  the tensor network simulations ({\it symbols}) with the single-meson prediction obtained for two different initial wave function: (i) the wave function $K(k)=-\I\tan(\theta^{\bar{h}_\perp}_k-\theta^{\hpe}_k)$ for the transverse field quench with no additional corrections ({\it dash-dot line});  (ii) the wave function ${\cal K}(k)=-\I\tan(\theta^{\bar{h}_\perp}_k-\theta^{\hpe}_k)-\I \hpa\bar{\sigma}v(k)/\epsilon^2(k) $ in Eq.~\eqref{WF}, obtained with a renormalized initial state ({\it full line}). The figure clearly shows that only a properly renormalized initial state $\ket{\tilde\psi_0}$ gives a good description of the entanglement dynamics.}\label{fig:state_prep}
\end{figure}

Finally, we can ri-exponentiate the state $\ket{\tilde{\psi}_0}$ to obtain the squeezed form
\be\label{squeezed}
|\tilde{\psi}_0\rangle\simeq\frac{1}{\sqrt{\mathcal{N}}}\exp\left[\int_0^\pi \dd k\,\mathcal{K}(k) \hat\gamma^\dagger(k)\hat\gamma^\dagger(-k)\right]|0\rangle\, .
\ee
Notice that Eq.~\eqref{squeezed} reduces to the linearized result above after a series expansion to the first order. Higher order terms in the squeezed form \eqref{squeezed} are justified from the physical assumption that pairs of fermions are incoherently created across the system. 
In Fig.~\ref{fig:state_prep}, we put to test our quantum prediction in Gaussian approximation obtained with the renormalized state \eqref{squeezed} and that for an initial state obtained from the wave function $K(k)=-\I\tan(\theta^{\bar{h}_\perp}_k-\theta^{\hpe}_k)$ of a transverse field quench. The comparison of the curves with tensor network simulations undoubtedly shows the relevance of the longitudinal field corrections on the initial state for a correct description of the entanglement dynamics.
\section{Computation of entanglement from Gaussian correlations}
\label{Ssec_EE}
By approximating  the time-evolved state $|\psi_t\rangle$ with a squeezed state, it is possible compute the  entanglement entropy with the help of well known formulas for free Fermi gases \cite{Casini2009}.
In order to do this, we first diagonalize the two-particle problem in the zero-momentum sector and subsequently we solve the Schr\"odinger equation \eqref{S_eq_Sh}. Since we explicitly used the translational invariance of the two-particle state, we found more convenient to work with a finite-size system having $N$ sites and periodic boundaries. Hence, we compute the correlation matrix in Eq.~\eqref{eq_C}
\be\label{S_eq_corr}
C=\int \frac{\dd k}{2\pi}\frac{e^{\I k(j-j')}}{1+|\mathcal{K}_t(k)|^2}U_{\theta_k} \begin{pmatrix} 1 && \mathcal{K}_t(k) \\ \mathcal{K}^*_t(k) && |\mathcal{K}_t(k)|^2  \end{pmatrix}U^\dagger_{\theta_k}\,,
\ee
where we used $\langle\hat\gamma^\dagger(q) \hat\gamma(k)\rangle=\delta(k-q) |\mathcal{K}(k)|^2/(1+|\mathcal{K}(k)|^2)$, $\langle\hat\gamma(q) \hat\gamma(-k)\rangle=\delta(k-q) \mathcal{K}(k)/(1+|\mathcal{K}(k)|^2)$ and we replaced the integration over momenta with a Fourier series.\\
It is convenient to organize the correlations in the matrix form
\be
C=
\begin{pmatrix} \mathbbm{1}-D &O\\
O^\dagger & D\end{pmatrix}\, ,\,\,\, D_{i,j}=\langle \hat{c}^\dagger_i \hat{c}_j\rangle \, ,\,\,\, O_{ij}=\langle \hat{c}_i \hat{c}_j\rangle
\ee
 from which the entanglement entropy can be conveniently computed. 
Following Ref.~\cite{Casini2009}, one notices that if the density matrix $\hat{\rho}$ is Gaussian, then the reduced density matrix $\hat{\rho}_A$ is also Gaussian and it is uniquely identified by the reduced correlation matrix $C_A$. 
For the sake of concreteness, let us assume the subsystem $A$ is an interval $j\in \{1,...,L\}$, but of course the method is more general.
As a next step, one considers a Bogoliubov rotation from the fermionic fields $\hat{c}_j$, $\hat{c}_j^\dagger$ to modes $\hat{f}_j$, $\hat{f}_j^\dagger$ that diagonalize the reduced density matrix
\be
\begin{pmatrix}\hat{c}_1\\ \hat{c}_2\\...\\ \hat{c}_L\\ \hat{c}^\dagger_1\\ \hat{c}^\dagger_2\\...\\ \hat{c}^\dagger_L \end{pmatrix}= U \begin{pmatrix}\hat{f}_1\\ \hat{f}_2\\...\\ \hat{f}_L\\ \hat{f}^\dagger_1\\ \hat{f}^\dagger_2\\...\\ \hat{f}^\dagger_L \end{pmatrix}
\ee
where $U$ is a $2L\times 2L$ matrix. In order to preserve the canonical commutation relations of the new fermionic modes, $U$ must be unitary. We fix $U$ by requiring that the correlation matrix $C_A$ is brought in the diagonal form
\be
U^\dagger C_A U= \begin{pmatrix} \mathbbm{1}-\Lambda &O\\
O^\dagger & \Lambda\end{pmatrix}
\ee
with $\Lambda$ a $L\times L$ diagonal matrix. Hence, $\hat{\rho}_A$ is also diagonal in the new fermionic basis 
\be\label{Seq_diaggauss}
\hat{\rho}_A\propto \exp\left(\sum_{i=1}^L\log\left(\frac{\Lambda_{ii}}{1-\Lambda_{ii}}\right)\hat{f}^\dagger_i \hat{f}_i \right).
\ee
From this result, the von Neumann and R\'enyi entropies of density matrices of the form \eqref{Seq_diaggauss} are easily obtained as
\be\begin{split}
&S_A=-\text{Tr}_A(\hat{\rho}_A\log\hat{\rho}_A)\\
&\quad= -\sum_i\Big\{ \Lambda_{ii} \log\Lambda_{ii}+ (1-\Lambda_{ii}) \log(1-\Lambda_{ii})\Big\} \\
&\quad=-\text{Tr}_A[C_A\log C_A]
\end{split}\ee
and
\be
S_A^{(n)}=\frac{1}{1-n}\log(\text{Tr}_A \  \hat{\rho}_A^n)= \frac{1}{1-n}\sum_i \log[\Lambda_{ii}^n+(1-\Lambda_{ii})^n].
\ee

Crucially, we assume periodic boundary conditions on the fermion basis when computing \eqref{S_eq_corr}, which imply a quantization of momenta as $k=2\pi n/N$. This introduces some subtleties in the computation of the Von Neumann and R\'enyi entropies, as we now discuss.
As already mentioned in Appendix~\ref{Ssec_ising_FF}, choosing periodic boundary conditions amounts to the choice of the parity sector with $P=-1$. 
Namely, the ground state correlation functions (obtained setting $\mathcal{K}=0$ in Eq.~\eqref{S_eq_corr}) are not those of $|{\rm GS}_\uparrow\rangle$, but rather those of the antisymmetric combination $|0^-\rangle=\frac{1}{\sqrt{2}}(|{\rm GS}_\uparrow\rangle-|{\rm GS}_\downarrow\rangle)$.
Before considering excited states and dynamics, it is instructive to compare the entanglement entropy obtained from the states $|{\rm GS}_\uparrow\rangle$ and $|0^-\rangle$.
More precisely, let us consider a bipartition $A\cup B$ where $A$ is an interval of length $L<N$ and let $S^{-}_L$ ( resp. $S^{\uparrow}_L$) be the entanglement entropy (which can be either von Neumann or R\'enyi, the forthcoming argument holds in both cases) of such a bipartition on the state $|0^-\rangle$ (resp. $|{\rm GS}_\uparrow\rangle$). With simple symmetry arguments between up and down spins, it is then easy to show that 
\be\label{S_eq_log}
S^{-}_L= S^{\uparrow}_L+\log 2\, .
\ee
As a simple check, one can set $\hpe=0$ and obtain Eq.~\eqref{S_eq_log} from the two degenerated ground states, which now take the form of product states with all spins up or down.
As a next step, we consider the excitations and their dynamics.
It is important to stress that, depending on the initial choice of the ground state ($|{\rm GS}_{\uparrow}\rangle$ or $|{\rm GS}_{\downarrow}\rangle$), one obtains very different excitations and, consequently, a different dynamics. However, by noticing that the Schr\"odinger equation \eqref{S_eq_Sh} describes the dynamics of excitations over the state $|{\rm GS}_{\uparrow}\rangle$, it is easy to see that using $\mathcal{K}_t$ in Eq.~\eqref{S_eq_corr}, we are not evolving $|0^-\rangle$, but we are rather separately evolving $|{\rm GS}_{\uparrow}\rangle$ with the Hamiltonian \eqref{eq_Ising} and $|{\rm GS}_{\downarrow}\rangle$ with a flipped sign in the longitudinal field $\hpa \to -\hpa$.
Therefore, the identity Eq.~\eqref{S_eq_log} remains valid also at finite times.
We then use the methods explained in Sec.~\ref{sec:Quantum} to compute $S^{-}_A$ at all times from the correlation matrix in Eq.~\eqref{S_eq_corr}, and we obtain $S^{\uparrow}_A$ by subtracting the constant offset $\log 2$. At this point, we have the entanglement entropy of a finite interval $L$ embedded in a system with periodic boundary conditions. On the other hand, we performed tensor network simulations to compute the half-size entanglement entropy of a finite system with open boundary conditions. To compare the two, we notice that the entanglement entropy obeys the area law, from which it follows that
\be
S_\text{half-sys; OBC}=\frac{1}{2} \lim_{L\to\infty} S_L^{\uparrow}\, .
\ee
In practice, saturation is obtained as soon as $L$ exceeds the maximum meson size $d_\text{max}= 4\hpe/(|\hpa|\bar{\sigma})$.

\section{Confinement in the long-range Ising chain}
\label{Ssec_long_range}
The dynamics of entanglement in the long-range Ising chain in Eq.~\eqref{eq_longH} can be obtained with the same strategy employed for the tilted Ising chain in Eq.~\eqref{eq_Ising}. 
Following Ref.~\cite{PhysRevLett.122.150601}, we focus on the regime of weak transverse field $\hpe\ll 1$, where the excitations of the model \eqref{eq_longH} are identified by domain walls. Restricting to the two domain walls subspace, the effective Hamiltonian is given by
\be\begin{split}
\hat{H}_\text{eff}|j_1,j_2\rangle&=-\hpe \Big[|j_1+1,j_2\rangle+|j_1-1,j_2\rangle+|j_1,j_2+1\rangle\\[4pt]
&+|j_1,j_2-1\rangle\Big]+U(|j_1-j_2|)|j_1,j_2\rangle\, ,
\end{split}\ee
where $|j_1,j_2\rangle\equiv|\uparrow...\uparrow_{j_1}\downarrow..\downarrow_{j_2-1}\uparrow_{j_2}...\rangle$ and, from hereafter, we consider an ordered set of coordinates with hard core constraint $j_1<j_2$ for the wave function. The two-kink interaction is
\be
U(j)= 4j \zeta(\alpha)-4 \sum_{1\le \ell<n}\sum_{1\le r\le \ell}\frac{1}{r^\alpha}\, ,
\ee
with $\zeta(\alpha)$ the Riemann-zeta function. Notice that the potential $U(j)$ diverges at large distances whenever $1<\alpha<2$. We are now interested to the dynamics of a ferromagnetic state with Hamiltonian \eqref{eq_longH} in the presence of a small, but finite, transverse field $\hpe$. Equivalently, we consider a small homogeneous quench in the transverse field. 
Therefore, one first considers the dynamics in the zero-momentum sector by defining $|j\rangle= \sum_{j'} |j,j+j'\rangle$. Similarly to the case of the Ising chain in a tilded field, we introduce the wave function $W_t(j)$ of two domain walls in the zero-momentum sector, satisfying the Schr\"odinger equation
\be\label{S_eq_Wt}
\I\partial_t W_t(j)= -2 \hpe\big[ W_t(j+1)+W_t(j-1)\big]+U(j)W_t(j)\, .
\ee
In the last expression, one should restrict to $j>0$ with the condition $W_t(j=0)=0$, due to the aforementioned choice of coordinates. In particular, a convenient way to automatically encode the hard core constraint is to consider an antisymmetric extension of the wave function to negative coordinates, i.e. imposing that $W_t(-j)=-W_t(j)$. When performing this operation, the potential is then symmetrically extended as $U(-j)=U(j)$. In other words, we are building a fermionic extension of the original wave function that is very useful for our scopes, as we now discuss.
Let $\Phi_n(j)$ be the normalized eigenfunctions in the antisymmetric sector of energy $E_n$, thus satisfying the time-independent Schr\"odinger equation
\be\label{S_eq_Sh_ind}
E_n \Phi_n(t)= -2 \hpe\big[ \Phi_n(j+1)+\Phi_n(j-1)\big]+U(j)\Phi_n(j)\, .
\ee
For fixed initial conditions, the wave function evolves as $W_t(j)=\sum_n e^{-\I E_n t} c_n \Phi_n(j)$ with time independent coefficients $c_n$. However, in the case at hand, such coefficients become time dependent $c_n\to c_n(t)$ since domain walls are \emph{dynamically} created from the fully-polarised state due to the transverse field. To compute $c_n(t)$, we proceed similarly to the short range Ising case and split the Hamiltonian \eqref{eq_longH} as $\hat{H}=\hat{H}_\text{diag}+\hat{H}_\text{off-diag}$, where $\hat{H}_\text{off-diag}$ is obtained by subtracting from $-\hpe \sum_i \hat{\sigma}_i^x$ the projected component acting within the sector with a conserved number of domain walls.
Then, we move to the interaction picture and consider first order perturbation theory by computing the matrix element of $-\I\int_0^{t} \dd \tau \, \hat{H}^{(I)}_{\text{off-diag}}(\tau)$ between the polarised state $|...\uparrow\uparrow\uparrow...\rangle$ and the eigenfunctions $\Phi_n$, reaching the simple expression
\be
W_t(j)=\sum_n e^{-\I E_n t} c_n(t) \Phi_n(j)
\ee
with coefficients
\be
c_n(t)= \hpe \frac{1-e^{\I t E_n}}{E_n}\sqrt{2}\Phi_n(1)\, .
\ee
The energies $E_n$ and eigenfunctions $\Phi_n$ are obtained numerically by solving Eq.~\eqref{S_eq_Sh_ind}. 
As a last step, we now wish to build the many-body wave function. In particular, we look for an effective description in terms of fermionic Gaussian states such that the formulas of Appendix \ref{Ssec_EE} for the entanglement remain still valid. However, in the long range setting, this requires to consider fictitious fermionic degrees of freedom instead of the standard fermionic excitations of the model (i.e., those obtainable through a Jordan-Wigner transformation). More precisely, for small transverse field $|h_\perp|\ll 1$, we notice that the fundamental excitations of the long range model are domain walls, similarly to those arising in the short range Ising model discussed previously. Therefore, one can introduce fictitious fermionic degrees of freedom for the long range problem as the domain wall excitations in the short range Ising model, and use these fictitious fermionic degrees of freedom to apply the strategy of Appendix \ref{Ssec_EE} for the calculation of the entanglement. Following this program, we write the two-body state in momentum space as a squeezed state of mode operators $\hat{\gamma}(k)$
\be
|\tilde{\psi}_0\rangle\propto\exp\left[\int_0^\pi \dd k\,  \mathcal{K}(k) \hat\gamma^\dagger(k) \hat\gamma^\dagger(-k)\right] |0\rangle\, ,
\ee
where $\mathcal{K}(k)$ is defined as the Fourier transform of the real space wave function $W_t(j)$.
Finally, the real space correlation functions of the fictitious fermions must be computed. To this end, one can simply use the expression in Eq.~\eqref{S_eq_corr} for the Ising model with the Bogoliubov angle $\theta_k=-\frac{1}{2\I}\log(-e^{\I k})$, which is nothing else than the limit $\hpe\to 0$ of Eq.~\eqref{Seq_Bog_angle}. From this result, one can obtain the entanglement entropy with the techniques of Appendix~\ref{Ssec_EE}.

\section{Breakdown of the single-meson description}\label{app:breakdown}
\begin{figure}[t]
\centering
\includegraphics[width=\columnwidth]{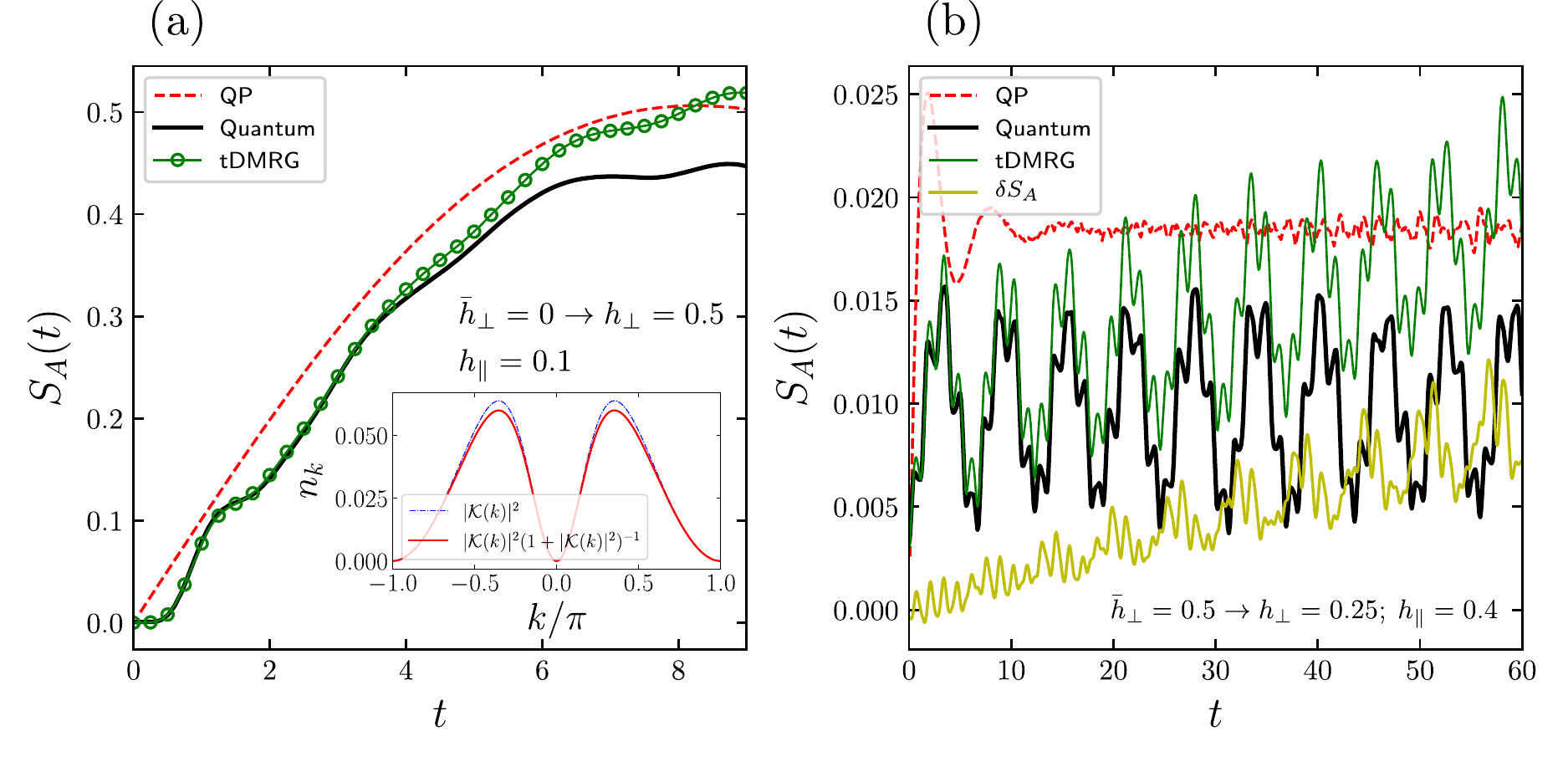}
\caption{(a) Large quench in the transverse field: $\bar{h}_\perp=0\to\hpe=0.5$, at $\hpa=0.1$. Inset: exact density of quasiparticles $n_k$ ({\it full line}) significantly deviates from its approximation ({\it dashed-dot line}) in the dilute regime. In this regime, both analytical predictions for the Von Neumann entropy (quasiparticle picture -- {\it dashed line}; quantum Gaussian treatment -- {\it full line}) fail to predict the observed numerical behavior ({\it symbols}) at finite times. (b) Example of Schwinger mechanism: $\bar{h}_\perp=0.5\to \hpe=0.25$ and strong confinement $\hpa=0.4$. The analytical predictions (quasiparticle picture -- {\it dashed line}; quantum Gaussian treatment -- {\it thick full line}) significantly deviates from the numerical data ({\it thin full line}). The difference $\delta S_A$ between the numerical data and the quantum prediction clearly shows a linear growth of entanglement due to the spontaneous creation/annihilation of new particles.}\label{fig:breakdown}
\end{figure}
\begin{figure}[t]
\centering
\includegraphics[width=\columnwidth]{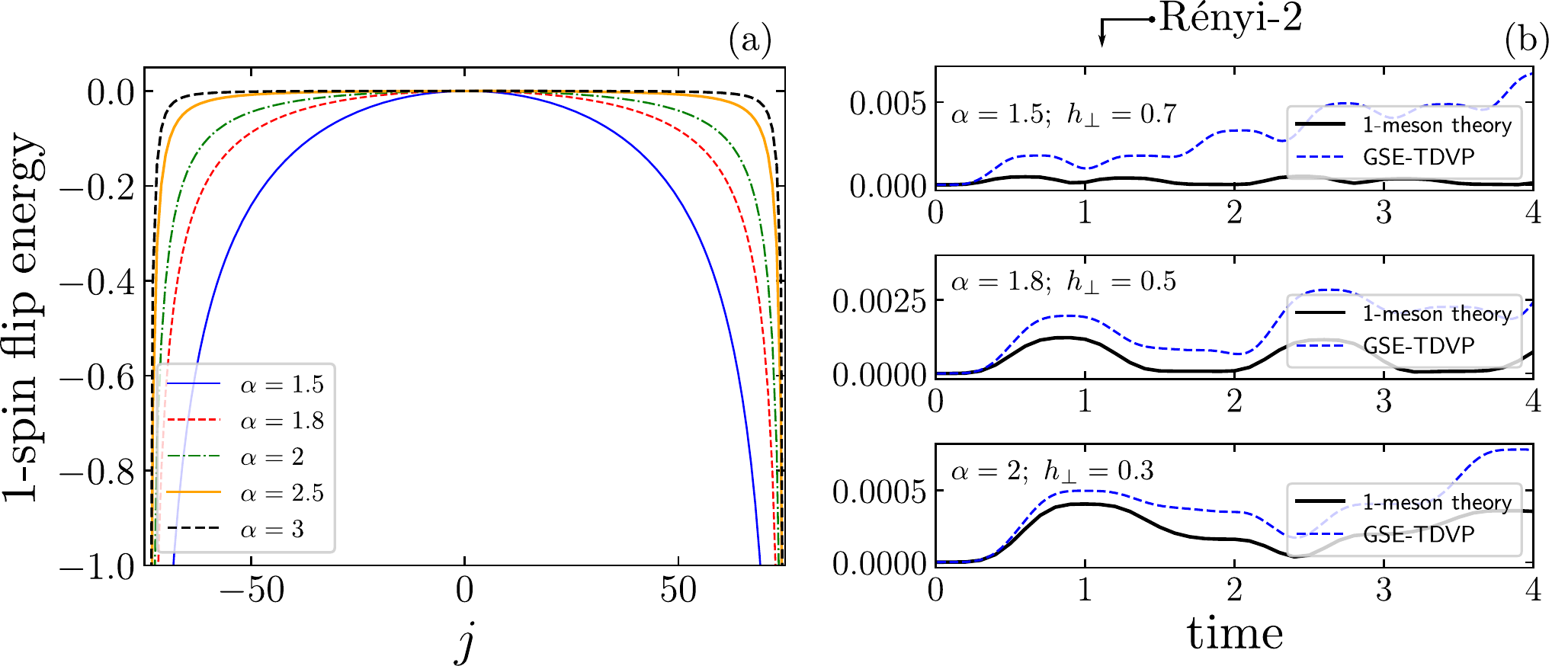}
\caption{(a) Single spin flip energy profile of a finite-size chain with $N=150$ sites for different values of $\alpha$. For $\alpha \lesssim 2$, the profile significantly deviates from a flat one and leads to a non-negligible boundary-meson interaction term. In each of the curves, we subtract the max value of the energy for a better comparison. (b) R\'enyi-2 entropy dynamics for the long-range Ising model (see Eq.~\eqref{eq_longH}) with exponent $\alpha\leq 2$ obtained with variational methods ({\it dashed line}). The emergence of meson-meson and meson-boundary interactions produces a linear drift which is not captured by the quantum prediction in single-meson approximation ({\it full line}). In each of the right panels, we set the transverse field $\hpe$ such that $\hpe/U(1)\ll 1$.
  }\label{fig:breakdown-long}
\end{figure}

In Sec.~\ref{sec:entanglement}, we derived an analytical prediction for the entanglement dynamics of confined spin chains under the assumption of small quenches, which physically corresponds to mesons that are well-separated and thus noninteracting.
This approximation is found in excellent agreement with the numerical data for a wide set of quench parameters, see Fig.~\ref{fig:results}. 
Nevertheless, corrections appear when the typical distance among mesons becomes comparable with their size or with the interaction range, as we now discuss.
To begin with, we consider such deviations in the short-range Ising model with tilted magnetic field (see Eq.~\eqref{eq_Ising}), where the assumption of well-separated mesons is easier to fulfill than in the long range case.
For this model, inter-mesons interactions are possible only through contact, therefore corrections are present only when the typical size of the meson, here estimated with the semiclassical maximum size $d_\text{max}=4\hpe/(\hpa\bar{\sigma})$, becomes comparable with the average spacing, the latter being the inverse of the meson density $n_\text{tot}=\int_0^\pi \frac{\dd k}{2\pi} \, n_k$. In Fig. \ref{fig:breakdown}~(a) we consider a large quench with $d_\text{max} n_\text{tot}\simeq 0.3$.
Semiclassically, pairs of fermions are originated at the same point in space, hence the size of mesons starts from zero and increases with time. Consequently, we observe an initial agreement with great accuracy between the tensor network data and the single-meson prediction, but the two drift away for $t>t^*$, with $t^*\simeq 4$ with this choice of parameters.
The regime of dilute meson can be better attained by either diminishing the density $n_\text{tot}$ or either reducing the size of mesons $d_\text{max}$ by increasing the longitudinal field. However, a too strong longitudinal field leads as well to a breakdown the single-meson approximation due to the Schwinger effect, through which the initial mesons decay in couples of lighter mesons at a constant rate.
In Fig.~\ref{fig:breakdown}~(b), we probed such quasiparticles production by considering a quench with strong confinement. We observe a linear growth of the deviation from the stable meson approximation. \\
We now refer to Fig.~\ref{fig:breakdown-long} for the long range case, where finite-size corrections to the single-meson predictions are particularly visible for small exponents $\alpha\lesssim 2$. As an example, in Fig.~\ref{fig:breakdown-long}~(a), we consider the energy of an isolated spin flip in a system of finite length $L=150$. The finite size of the system induces a non trivial potential landscape and the meson (or, more precisely, the deep boundstate for $\alpha>2$) is initially created at rest and later accelerated due to the effective inhomogeneous potential. This causes a scattering of the excitations which results in a linear growth of entanglement on intermediate time scales, see Fig.~\ref{fig:breakdown-long}~(b). Such finite size effects are suppressed at larger $\alpha$, see Fig.~\ref{fig:long_range}.\\ 
Although, strictly speaking, $\alpha>2$ does not induce confining dynamics, we experienced that  setting $2.5\leq\alpha\leq 3$ is a good compromise between having deep bound states and relatively small boundary effects on the sizes we analyzed.
Furthermore, we notice that, even in the ideal scenario of an infinite system, similar long-tailed interactions will show up among mesons (or bound states). Hence, we conclude that the dilute regime is attained when the distance among the excitations is large compared to the $\alpha-$dependent interaction range. In Fig.~\ref{fig:breakdown-long} ~(b) we consider the entanglement growth after transverse field quenches for values of $\alpha$ smaller than those considered in Sec.~\ref{sec:long-range}. In particular, corrections with respect to our predictions are more relevant as $\alpha$ is reduced, in agreement with the phenomenology we depicted.

\bibliography{biblio}
\end{document}